\documentclass[aoas,MSNbibl,nameyear,noautosecdot,dvips]{arximspdf}
\usepackage{graphicx,breakurl}
\doi{10.1214/13-AOAS614D} 
\referstodoi{10.1214/12-AOAS614}
\volume{7}
\issue{4}
\pubyear{2013}
\firstpage{1881}
\lastpage{1887}

\makeatletter
\newcommand{\etal}{et al.}
\newcommand{\X}{\mathbf{X}}
\newcommand{\range}{\dvtx}
\newcommand{\B}{{\bolds{\beta}}}
\makeatother

\begin{document}
\begin{frontmatter}

\title{Discussion of ``Estimating the historical and future
probabilities of large terrorist events'' by Aaron Clauset and Ryan Woodard}
\runtitle{Discussion}

\begin{aug}
\author[A]{\fnms{Jeff} \snm{Gill}\corref{}\ead[label=e1]{jgill@wustl.edu}}
\runauthor{J. Gill}
\affiliation{Washington University in St. Louis}
\address[A]{Department of Political Science\\
Division of Biostatistics\\
Department of Surgery (Public Health Sciences)\\
{Washington University in St. Louis}\\
One Brookings Drive, Campus Box 1063\\
Saint Louis, Missouri 63130\\
USA\\
\printead{e1}} 
\pdftitle{Discussion of ``Estimating the historical and future
probabilities of large terrorist events'' by Aaron Clauset and Ryan Woodard}
\end{aug}

\received{\smonth{7} \syear{2013}}



\end{frontmatter}

\section{\texorpdfstring{Introductory remarks.}{Introductory remarks}}\label{sec1}
In developing a new algorithm for estimating probabilities for tail
events Clauset and Woodard have provided an important new tool
for understanding social events that are rare and momentous. Such
upper-tail large-scale events are notoriously hard to predict
because there is obviously less certainty relative to more typical
events. So even unbiased estimates, which are difficult, are
likely to have large confidence intervals. This is further exacerbated
by the measurement error challenges inherent in nearly all
aggregated social science data.

The safety of millions of people depends on understanding the
intentions and actions of terrorist groups. To protect
citizens, governments and nongovernmental organizations invest enormous
amounts of time and energy attempting to detect malevolent
covert groups and to thwart terrorist attacks. These terrorist events
vary dramatically in scope, but are usually measured in
terms of casualties (injuries and fatalities). However, the effect of
terrorist attacks can be quite substantial even with modest
casualties; the Boston Marathon bombing of 2013 had ``only'' three
fatalities and yet had a great effect on the nation's psyche.
So it is considered to be a successful terrorist event by observers and
scholars of terrorism. Why is this? It is because the
real intention of a terrorist is not just to kill people; this is an
intermediate step. The real intention of terrorists is to
make citizens feel that their government cannot protect them. This is
designed to create unrest and lead to a change of
government policies in a direction favored by the terrorists, or a
failure of government, presumably to be replaced with one that
is preferred by the terrorists. Therefore, the more grisly (blood,
gore, beheadings, hanging bodies, etc.) and the more seemingly
random the victims, the greater the psycho-social effect on the
population. Examples are unfortunately plentiful throughout the
world: the Taliban wants to replace the current US-supported Afghan
government, the Moro Island Liberation Front wants to topple
the government in Manila to create a separate Islamist government in
the South Philippine islands, Al-Qaeda in the Islamic Maghreb
wants to overthrow the government of Algeria and neighboring countries
to form an Islamic state in Northern Africa, the
Revolutionary Armed Forces of Colombia (FARC) is still active in trying
to destabilize the current government and replace it with
a Marxist---Leninist alternative. The US State Department actually
lists 51 active ``official'' Foreign Terrorist Organizations.

The quantity and quality of events data to understand this problem
(where the unit of analysis is a single attack) have improved
dramatically, but are still poor relative to other social science data.
Modeling with these data has yielded some information
about the determinants and timing of terrorist incidents [\citet{End07}, \citet{EndSan95} and \citet{LiSch04}, for
example]. However, our ability to empirically model terrorist
activities currently is limited because these data consist mostly
of observed and recorded terrorist events only. Exceptions include Ed
Mickolus' (\citeyear{Mic82}) ITERATE biographical data set of terrorists. So
these data do not constitute the complete set of the activities of
these actors since attacks may get canceled or altered,
governments are sometimes motivated not to report thwarted activities.
This dependence on events-only data violates the standard
admonition in the social sciences of not selecting on the outcome
variable [\citet{KinKeoVer94}]. Measurement issues
are also often a serious problem: the estimated number of casualties
for a single attack can be uncertain, the attacking group may
not be obvious, eyewitnesses can vary in their description, terrorists
are motivated to hide processes, methods and
capabilities. However, researchers have little choice but to contend
with such data challenges. I have personally confronted
these methodological issues [Gill and Freeman (\citeyear{GF13}), Kyung \etal\ (\citeyear{KyuGilCas11N1,KyuGilCas11N2})].

So into this literature we have a new contribution. Clauset and Woodard
provide a novel method for understanding an important
feature of terrorist data: what is the probability of a catastrophic
large-scale event over some period of time? Their paper
provides a new and highly-valuable tool for assessing risk based on an
empirical distribution of known events. This will enable
academic and government analysts to effectively assess, and perhaps
plan for, extremely large (e.g., successful) attacks. Their
paper is a major contribution to this substantive area. Furthermore,
this work is a classic contribution to the \emph{Annals of
Applied Statistics} in the sense that it combines a critical real-world
problem with a new statistical method to produce new
insights.\looseness=-1

The authors have cleverly combined a threshold specification and
alternative parametric model comparison, with nonparametric
bootstrapping. The threshold here, $x_{\mathrm{min}}$, is simply a value
that allows us to dispense with the left-hand side of some
PDF for modeling purposes. Thus, the right-hand tail only is modeled
over $[x_{\mathrm{min}}\range\infty)$, which gives added
flexibility by avoiding fitting the more common occurrences at the
same time. Obviously, this still leaves a wide range of
parametric specifications defined of this support with declining
density moving to the right, so Clauset and Woodard test common
alternatives with standard likelihood ratio tests. Unfortunately, this
is not enough since the parameters of these PDFs are
sensitive to instability in the empirical data over this region,
requiring another step whereby the models are weighted by their
likelihood from a (nonparametric) bootstrap distribution. This allows
them to construct extreme value confidence intervals from
standard theory.

The core of the approach is establishing $p_{\mathrm{tail}}$ as the
observed proportion of events equal to or larger than
$x_{\mathrm{min}}$ in each of the bootstrapped samples. Thus, $n_{\mathrm
{tail}}$ is simply a binomial outcome from $m$ bootstrap
samples with probability $p_{\mathrm{tail}}$. From an assumed
distribution, this leads to the probability of observing a
large-scale event (or events). The problem of course is the selection
of this distribution, and the authors compare the discrete
power-law distribution to the log-normal and the stretched exponential
distributions for this purpose. What fixed value of
$x_{\mathrm{min}}$ should be used? The lack of a theoretically driven
threshold suggests that an effective strategy would be to
estimate the starting point of the upper tail used. Unfortunately, the
authors' bootstrap models return about half of the
estimates of $\hat{x}_{\mathrm{min}}$ around 9--10 but with a large
proportion also at 4--5. Apparently 10 is a good value in that
continuous and discrete tail models produce similar up tail structures,
and this value is used throughout most of the empirical
work, except where it is estimated (e.g., $\hat{x}_{\mathrm{min}} = 39$
in Section~C.2). An extension where estimation of
$x_{\mathrm{min}}$ is conditional on covariates, informed prior
distributions or other relevant information would be a welcome
addition to the existing model.

\section{\texorpdfstring{Discussion questions.}{Discussion questions}}\label{sec2}
This section discusses some important issues raised by Clauset and
Woodard. As noted, terrorism data is extremely difficult to
model and this section is not intended to diminish the progress made in
their paper.

\subsection{\texorpdfstring{Why focus on outlier events?}{Why focus on outlier events?}}\label{sec2.1}
Are bigger events in terms of the number of fatalities really the
``bigger'' events? Since the purpose of terrorism is to exert
psychosocial instability, more deaths might not be bigger events. The
key is distance and circumstance. Consider two events in
the same month of May 2013. On May 22 a single off-duty British soldier
in the Woolwich district in South East London was
run down by two assailants with their car and then brutally hacked
apart with knives and a machete. A week earlier a coordinated
series of attacks in Iraq killed 449 people. Nearly every citizen of
westernized countries (and more) immediately knew about the
May 22 event, and only a small fraction paid attention to the earlier
event, \emph{despite the fact that it was 449 times more\vadjust{\goodbreak}
deadly}. Obviously, the London attack spread more ``terror'' because it
was closer to supposedly safe citizens and because Iraq is
still perceived as a distant war zone by many. Since all major
terrorist attacks result in psychiatric morbidity for some of the
population [\citet{Cri04}], the question is whether in the context of
the attack (place, casualties, damage, media coverage) the
number killed is always the most important factor. Certainly this is
not true.

\subsection{\texorpdfstring{Is it I.I.D.?}{Is it I.I.D.?}}\label{sec2.2}
The finding that $\hat{x}_{\mathrm{min}}$ is bimodal when estimated in
the context of the bootstrap models suggests that there may
be two or more eras of terrorism in the data. The RAND-MIPT data used
covers 1968 to 2007, which is a long period to assume that
terrorism is stable and consistent in strategy, effectiveness and
methods. Furthermore, RAND-MIPT data is based on a very broad
definition of what constitutes a terrorist event, where some are better
labeled as war crimes. These types of data-labeling
distinctions are a major reason why different terrorism data sets
report different events. The authors discuss the i.i.d. issue in the
fourth paragraph of page~16, stating that the ``i.i.d. assumption appears
to be statistically justified at the global spatial and
long-term temporal scales studied here.'' But this is clearly not the
case empirically, as major home-grown terrorism in Western
Europe has declined dramatically since the demise of the PIRA,
Baader-Meinhof and other groups. Terrorism was virtually unknown
in Eastern Europe before the collapse of the Soviet Union, but now
Chechen and Chechen-inspired terror is a regular (and now
exported) phenomenon. India has lately emerged as a major attractor of
terrorism. Also, during the cold war era major powers
tended to suppress the definition of terrorism if it suited their
purposes. For instance, the Contras in Nicaragua were never
considered by the US to be terrorists (despite the opposite finding by
the International Court of Justice), even though their
alleged acts fall under the RAND-MIPS definitions.

%

Fortunately or unfortunately, there is not a single definitive data set
for terrorism events. In addition to RAND-MIPT, frequently
used alternatives include the Global Terrorism Database at the
University of Maryland (describing over 104,000 attacks from 1970
to 2011), ITERATE, the Big, Allied and Dangerous (BAAD) Database~1
[Asal, Rethemeyer and Anderson (\citeyear{AsaKarAnd}), aggregating worldwide
lethal attacks from 1998--2005], the Worldwide Incidents Tracking System
(WITS) from the National Counterterrorism Center starting
in 2004, data sets collected by government agencies and more. All of
these show various trends over time, and countless articles
have been written about eras of terrorism. For example, Kyung \etal\ modeled suicide attack events data in the Middle East and
Northern Africa from 1998 to 2004 using a Dirichlet process random
effects model and found that 1998 was an exception year that
could not be considered as coming from the same distribution as the
other years in the study (there were 273 major terrorist
attacks worldwide in 1998 with an astonishing 741 killed and 5952
injured).\eject

\begin{figure}

\includegraphics{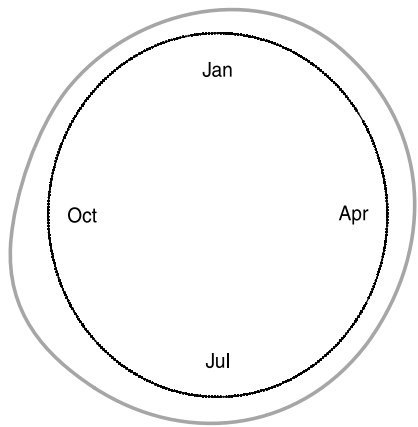}

\caption{GTD2 yearly cycle.}\label{Circ.Fig}
\end{figure}

Consider also incidents from the Global Terrorism Database II [\citet{LaFDug08}] as shown to the right. The plot
gives a kernel density for the number of killed by day of the year
across the years 1998--2004 with the 9/11 attacks removed for
scale purposes [see Gill and Hangartner (\citeyear{GilHan10}) for details of circular
analysis for social science data]. Clearly the data show a
nonuniform pattern by time with a particular bulge around September. So
for this short 7-year period there is both a yearly trend
(1998 and 2001 are exceptional) and a within-year trend. Enlarging the
time period makes this effect worse because national and
international trends undoubtedly \emph{add} more heterogeneity. This
issue is addressed in the authors' Sections~3 with a discussion
of covariates. The authors rerun the tail models under different
circumstances as a means of controlling for the following: different time
periods, same/different country for attacker/target, country economic
status and type of weapon used. Instead of separate
models, it would be more satisfying to see a GLM-style development,
which would be easier with the provided log-normal
specification since $\mu= \X\B$ is a natural parameterization in
that context.

\subsection{\texorpdfstring{Why not be Bayesian about this?}{Why not be Bayesian about this?}}\label{sec2.3}
Clauset and Woodard state that ``a~Bayesian approach would be
inconsistent with our existing framework'' (page~6). This may not
be necessarily true. Recall the ability in Bayesian inference for
serial updates as new information arrives. That is, posterior
distributions for parameters of interest that were produced from data
and a prior distribution can serve as priors for the same
process in a future period as new data arrives. The resulting
second-stage posterior is the same form as if both sets of data had
arrived at the same time. This learning process could be used to update
the parameters of the tail models specified in the paper.
For instance, Table~1 shows that the log-normal tail model performs
poorly relative to the alternatives. However, if $\mu$ and
$\sigma$ in $p(x|\mu,\sigma) \propto x^{-1}\exp [ -(\log x -
\mu)^2/2\sigma^2  ]$ were updated over time (and there is
plenty of time in these data), it may outperform less parametrically
flexible alternatives.

\begin{figure}

\includegraphics{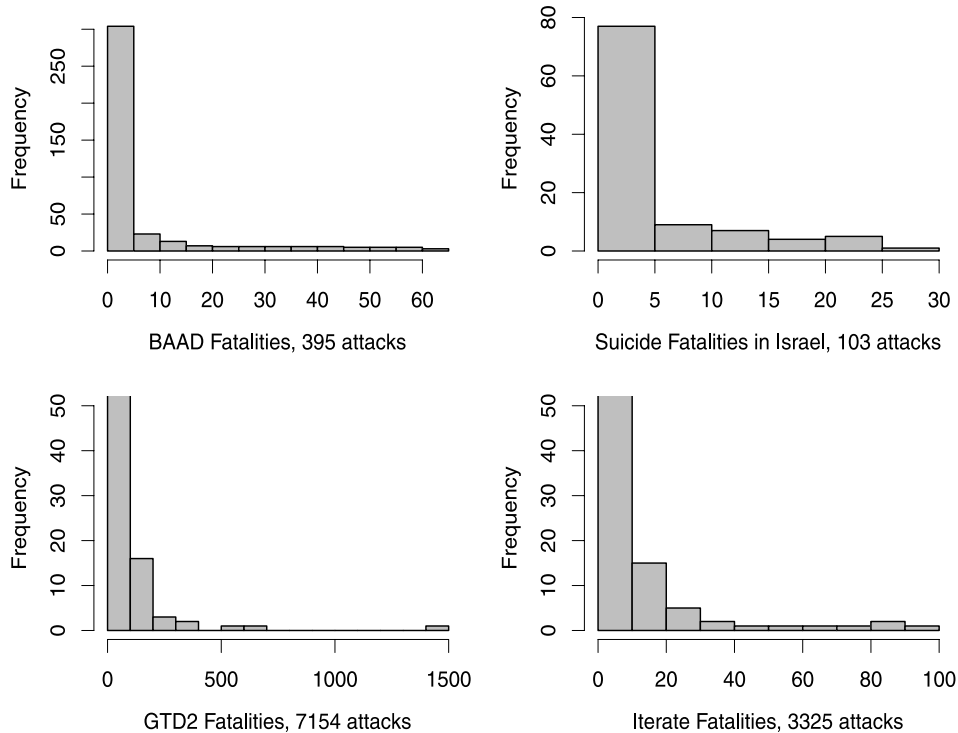}

\caption{Histogram of total killed by data set.}\label{Hist.Fig}
\end{figure}

\subsection{\texorpdfstring{Distributional forms.}{Distributional forms}}\label{sec2.4}
The requirement of a specific PDF for the tail model here is a big
convenience. The question is whether any of the alternatives
tried here, or others, are appropriate for these kinds of tails.
Consider the histograms in Figure~\ref{Hist.Fig} that show: (1)
fatalities in the BAAD data for allied terrorist groups 1998--2005, (2)
suicide attacks in Israel in the early period of the first
``Intifada'' November 6, 2000 to November 3, 2003 [see Kyung \etal\ (\citeyear{KyuGilCas11N1}) for details], (3) the Global Terrorism Database II
fatalities discussed above, and (4)~fatalities from the ITERATE data
set 1968--1977. The $y$-axis of the last two is truncated in
order not to obscure the distribution of the tail values with a large
range. Obviously these are only a small set of
examples, but it is clear from the heterogeneity of forms that a single
parametric tail model would not be appropriate in all of
these circumstances. This issue is briefly noted in the authors'
Section~5, and is obviously related to the i.i.d. above discussion.

\section{\texorpdfstring{Final words.}{Final words}}\label{sec3}
I congratulate Clauset and Woodard for taking on a difficult problem
and making substantial progress. This is an excellent
contribution to two literatures. My concerns above are mild and
primarily reflect the difficulty in dealing with data that comes
from a complex social and political process with violent actors
attempting to hide information about their characteristics. The
heterogeneity in methods, tools, geography and successes is also not
helpful to the statistical modeler. Despite these
challenges, we have learned something about the occurrences of extreme
terrorist events over time from this excellent paper.




\printaddresses


\begin{thebibliography}{12}

\bibitem[\protect\citeauthoryear{Asal, Karl~Rethemeyer and
Anderson}{2009}]{AsaKarAnd}
\begin{bmisc}[auto:STB|2013/10/09|12:32:29]
\bauthor{\bsnm{Asal},~\bfnm{Victor}\binits{V.}},
\bauthor{\bsnm{Karl~Rethemeyer},~\bfnm{R.}\binits{R.}} \AND
\bauthor{\bsnm{Anderson},~\bfnm{Ian}\binits{I.}}
(\byear{2009}).
\bhowpublished{Big allied and dangerous (BAAD) Database 1---Lethality
data, 1998--2005.
Available at \url{http://www.start.umd.edu/start/announcements/announcement.asp?id=357}}.
\bptok{imsref}%
\end{bmisc}
\endbibitem

\bibitem[\protect\citeauthoryear{Crimando}{2004}]{Cri04}
\begin{barticle}[auto:STB|2013/10/09|12:32:29]
\bauthor{\bsnm{Crimando},~\bfnm{Steven~M.}\binits{S.~M.}}
(\byear{2004}).
\btitle{The bio-psycho-social consequences of terrorism}.
\bjournal{Public Health Emergencies: Terrorism Preparedness}
\bvolume{101}
\bpages{84--89}.
\bptok{imsref}%
\end{barticle}
\endbibitem

\bibitem[\protect\citeauthoryear{Enders}{2007}]{End07}
\begin{bincollection}[auto:STB|2013/10/09|12:32:29]
\bauthor{\bsnm{Enders},~\bfnm{Walter}\binits{W.}}
(\byear{2007}).
\btitle{Terrorism: An empirical analysis}.
In \bbooktitle{Handbook of Defense Economics, Vol. 2}
(\beditor{\binits{T.}\bfnm{T.} \bsnm{Sandler}}
\AND
\beditor{\binits{K.}\bfnm{K.} \bsnm{Hartley}}, eds.).
\bpublisher{Elsevier}, \baddress{Amsterdam}.
\bptok{imsref}%
\end{bincollection}
\endbibitem

\bibitem[\protect\citeauthoryear{Enders and Sandler}{1995}]{EndSan95}
\begin{bincollection}[auto:STB|2013/10/09|12:32:29]
\bauthor{\bsnm{Enders},~\bfnm{Walter}\binits{W.}} \AND
\bauthor{\bsnm{Sandler},~\bfnm{Todd}\binits{T.}}
(\byear{1995}).
\btitle{Terrorism: Theory and applications}.
In \bbooktitle{Handbook of Defense Economics, Vol. 1}
(\beditor{\binits{K.}\bfnm{K.} \bsnm{Hartley}}
\AND
\beditor{\binits{T.}\bfnm{T.} \bsnm{Sandler}}, eds.).
\bpublisher{Elsevier}, \baddress{Amsterdam}.
\bptok{imsref}%
\end{bincollection}
\endbibitem


\bibitem[\protect\citeauthoryear{Gill and Freeman}{2013}]{GF13}
\begin{barticle}[auto]
\bauthor{\bsnm{Gill},~\bfnm{Jeff}\binits{J.}} \AND
\bauthor{\bsnm{Freeman},~\bfnm{John}\binits{J.}}
(\byear{2013}).
\btitle{Dynamic elicited priors for updating covert networks}.
\bjournal{Network Science}
\bvolume{1}
\bpages{68--94}.
\bptok{imsref}%
\end{barticle}
\endbibitem

\bibitem[\protect\citeauthoryear{Gill and Hangartner}{2010}]{GilHan10}
\begin{barticle}[auto:STB|2013/10/09|12:32:29]
\bauthor{\bsnm{Gill},~\bfnm{Jeff}\binits{J.}} \AND
\bauthor{\bsnm{Hangartner},~\bfnm{Dominik}\binits{D.}}
(\byear{2010}).
\btitle{Circular data in political science and how to handle it}.
\bjournal{Political Analysis}
\bvolume{18}
\bpages{316--336}.
\bptok{imsref}%
\end{barticle}
\endbibitem

\bibitem[\protect\citeauthoryear{King, Keohane and Verba}{1994}]{KinKeoVer94}
\begin{bbook}[auto:STB|2013/10/09|12:32:29]
\bauthor{\bsnm{King},~\bfnm{Gary}\binits{G.}},
\bauthor{\bsnm{Keohane},~\bfnm{Robert~O.}\binits{R.~O.}} \AND
\bauthor{\bsnm{Verba},~\bfnm{Sidney}\binits{S.}}
(\byear{1994}).
\btitle{Designing Social Inquiry: Scientific Inference in Qualitative
Research}.
\bpublisher{Princeton Univ. Press}, \blocation{Princeton, NJ}.
\bptok{imsref}%
\end{bbook}
\endbibitem

\bibitem[\protect\citeauthoryear{Kyung, Gill and
Casella}{2011}]{KyuGilCas11N2}
\begin{barticle}[mr]
\bauthor{\bsnm{Kyung},~\bfnm{Minjung}\binits{M.}},
\bauthor{\bsnm{Gill},~\bfnm{Jeff}\binits{J.}} \AND
\bauthor{\bsnm{Casella},~\bfnm{George}\binits{G.}}
(\byear{2011}).
\btitle{Sampling schemes for generalized linear {D}irichlet process random
effects models}.
\bjournal{Stat. Methods Appl.}
\bvolume{20}
\bpages{259--290}.
\bid{doi={10.1007/s10260-011-0168-x}, issn={1618-2510}, mr={2859768}}
\bptnote{check year}%
\bptok{imsref}%
\end{barticle}
\endbibitem

\bibitem[\protect\citeauthoryear{Kyung, Gill and
Casella}{2012}]{KyuGilCas11N1}
\begin{barticle}[mr]
\bauthor{\bsnm{Kyung},~\bfnm{Minjung}\binits{M.}},
\bauthor{\bsnm{Gill},~\bfnm{Jeff}\binits{J.}} \AND
\bauthor{\bsnm{Casella},~\bfnm{George}\binits{G.}}
(\byear{2012}).
\btitle{New findings from terrorism data: {D}irichlet process random-effects
models for latent groups (with discussion and rejoinder)}.
\bjournal{J. R. Stat. Soc. Ser. C. Appl. Stat.}
\bvolume{60}
\bpages{701--721}.
\bid{doi={10.1111/j.1467-9876.2011.00768.x}, issn={0035-9254}, mr={2844851}}
\bptok{imsref}%
\end{barticle}
\endbibitem



\bibitem[\protect\citeauthoryear{LaFree and Dugan}{2008}]{LaFDug08}
\begin{bbook}[auto:STB|2013/10/09|12:32:29]
\bauthor{\bsnm{LaFree},~\bfnm{Gary}\binits{G.}} \AND
\bauthor{\bsnm{Dugan},~\bfnm{Laura}\binits{L.}}
(\byear{2008}).
\btitle{Global Terrorism Database II, 1998--2004 [Computer File].
ICPSR22600-v2}.
\bpublisher{Inter-university Consortium for Political and Social Research}, \blocation{Ann Arbor, MI}.
\bptok{imsref}%
\end{bbook}
\endbibitem

\bibitem[\protect\citeauthoryear{Li and Schaub}{2004}]{LiSch04}
\begin{barticle}[auto:STB|2013/10/09|12:32:29]
\bauthor{\bsnm{Li},~\bfnm{Quan}\binits{Q.}} \AND
\bauthor{\bsnm{Schaub},~\bfnm{Drew}\binits{D.}}
(\byear{2004}).
\btitle{Economic globalization and transnational terrorist incidents}.
\bjournal{Journal of Conflict Resolution}
\bvolume{48}
\bpages{230--258}.
\bptok{imsref}%
\end{barticle}
\endbibitem

\bibitem[\protect\citeauthoryear{Mickolus}{1982}]{Mic82}
\begin{bbook}[auto:STB|2013/10/09|12:32:29]
\bauthor{\bsnm{Mickolus},~\bfnm{Edward~F.}\binits{E.~F.}}
(\byear{1982}).
\btitle{International Terrorism: Attributes of Terrorist Events, 1968--1977
[ITERATE 2 Computer File]. ICPSR07947-v1}.
\bpublisher{Inter-university Consortium for Political and Social Research}, \blocation{Ann Arbor, MI}.
\bptok{imsref}%
\end{bbook}
\endbibitem

\end{thebibliography}
\end{document}